\documentclass[%
reprint,
superscriptaddress,
 amsmath,amssymb,
prfluids
longbibliography,
]{revtex4-2}

\usepackage{romannum}
\usepackage{graphicx}
\usepackage{dcolumn}

\usepackage{siunitx}
\usepackage{hyperref}

\usepackage{xcolor}
\definecolor{myred}{RGB}{255,3,0}
\definecolor{myblue}{RGB}{31,60,255}
\definecolor{myblue2}{RGB}{34,123,255}
\definecolor{myorange}{RGB}{226,48,0}
\definecolor{mygreen}{RGB}{21,100,0}
\definecolor{mygreen2}{RGB}{118,172,66}

\usepackage{tikz}
\usetikzlibrary{shapes.geometric,shapes.symbols}

\newcommand{\tikzcircle}[2][red,fill=red]{\tikz[baseline=-0.5ex]\draw[#1,radius=#2] (0,0) circle ;}
\newcommand{\tikzsymbol}[2][circle]{\tikz[baseline=-0.75ex]\node[inner
sep=2pt,shape=#1,draw,#2]{};}

\newcommand{\tikztriangle}[2][triangle]{\tikz[baseline=-0.3ex]\node[inner
sep=1.75pt,isosceles triangle,isosceles triangle apex angle=60,draw,#2,rotate=90]{};}

\begin{document}

\title{\large{Frozen Cheerios effect: Particle-particle interaction \\ induced by an advancing solidification front}}

\author{Jochem G. Meijer}
\thanks{These two authors contributed equally}
\email[]{jgmeijer@uchicago.edu}
\affiliation{Physics of Fluids Department, Max Planck Center Twente for Complex Fluid Dynamics, and J. M. Burgers Center for Fluid Dynamics, Faculty of Science and Technology, University of Twente, P.O. Box 217, 7500 AE Enschede, The Netherlands}

\author{Vincent Bertin}%
\thanks{These two authors contributed equally}
\email[]{vincent.bertin.1@univ-amu.fr}
\affiliation{Physics of Fluids group, Max Planck Center Twente for Complex Fluid Dynamics, Department of Science and Technology, Mesa+ Institute and J. M. Burgers Center for Fluid Dynamics, University of Twente, P.O. Box 217, 7500 AE Enschede, The Netherlands}

\author{Detlef Lohse}%
\email[]{d.lohse@utwente.nl}
\affiliation{Physics of Fluids group, Max Planck Center Twente for Complex Fluid Dynamics, Department of Science and Technology, Mesa+ Institute and J. M. Burgers Center for Fluid Dynamics, University of Twente, P.O. Box 217, 7500 AE Enschede, The Netherlands}
\affiliation{Max Planck Institute for Dynamics and Self-Organization, Am Fassberg 17, 37077 G\"ottingen, Germany}

\begin{abstract}
Particles at liquid interfaces have the tendency to cluster due to capillary forces competing with gravitational buoyancy (\textit{i.e.,} normal to the distorted free surface).
This is known as the Cheerios effect. 
Here we experimentally and theoretically study the interaction between two submerged particles near an advancing water-ice interface during the freezing process. 
Particles that are thermally more conductive than water are observed to attract each other and form clusters once frozen. 
We call this feature the frozen Cheerios effect, where interactions are driven by alterations to the direction of the experienced repelling force (\textit{i.e.}, normal to the distorted isotherm).
On the other hand, particles less conductive than water separate, highlighting the importance of thermal conduction during freezing. 
Based on existing models for single particle trapping in ice, we develop an understanding of multiple particle interaction.
We find that the overall efficacy of the particle-particle interaction critically depends on the solidification front velocity.
Our theory explains why the thermal conductivity mismatch between the particles and water dictates the attractive or repulsive nature of the particle-particle interaction.  
\end{abstract}

\date{\today}

\maketitle
\onecolumngrid
\section{Introduction}
The freezing of aqueous solutions containing rigid or soft particles (droplets, cells, bubbles, etc.) is ubiquitous in many natural and industrial settings~\cite{deville2017freezing}.  Typical examples are found in the food industry~\cite{amit2017review}, frost-heaving in cold regions \cite{rempel2010frost,peppin2011frost,peppin2013physics}, or the cryopreservation of biological tissue \cite{bronstein1981rejection,korber1988phenomena,muldrew2004water}.  During solidification, the immersed particles are subjected to stresses~\cite{gerber2022stress}, such that they may deform~\cite{carte1961air,maeno1967air,bari1974nucleation,wei2000shape,wei2002analytical,wei2004growths, tyagi2022solute,meijer2023thin,thievenaz2025on} or get displaced~\cite{korber1985interaction,lipp1993engulfment}. As a result, during the freezing of a randomly dispersed suspension, particles can accumulate at the freezing front upon freezing, which eventually leads to the formation of intricate particle clusters in the resulting solid~\cite{saint2017interaction,tyagi2021multiple}, which is crucial for templating directionally porous materials~\cite{deville2008freeze,deville2022complex}.

The interaction between a single particle with an advancing solidification front is a delicate interplay between intermolecular repulsive forces, thermal effects and viscous friction, all of which act on different length and time scales~\cite{shangguan1992analytical}. Typically, at low freezing speed, the repulsive force dominates and the particles are always segregated from the front and pushed away by it. Conversely, at large freezing speed, the viscous friction is large and drives the engulfment of the particle in the newly-formed solid. A large body of work has been devoted to the determination of the critical front speed for engulfment, either theoretically with asymptotic matching techniques~\cite{rempel1999interaction,rempel2001particle,park2006encapsulation} or numerically~\cite{garvin2007multiscale,tao2016steady}. In this article, we focus on the limiting case where the front speed is close to the critical engulfment speed. 

In this situation, particles are significantly displaced by the front before possibly getting engulfed~\cite{dedovets2018five}. Hence, other particles in the suspension may also enter in contact with the freezing front at the same time, leading to collective interactions between multiple particles and the freezing front~\cite{tyagi2021multiple}. These multi-body interactions are very complex, even for monodisperse suspensions. For instance, particles have been observed to get engulfed collectively, eventually forming a line of particles oriented in the front direction. Additionally, in polydisperse suspensions, large particles have been observed to entrain smaller particles in their wake, evidencing hydrodynamic interactions~\cite{tyagi2021multiple}. While the single particle-front interaction is fairly well understood, the understanding of the interaction between accumulated particles at the freezing front is currently lacking. In this article, we focus on the simplest possible multi-body interactions between particles and freezing front which is the two-body particle-particle interactions. 

We have performed model experiments by freezing different suspensions in a unidirectional manner and focus the analysis on the binary interactions. The main experimental finding is that the attractive/repulsive nature of the particle-particle interaction depends on the relative thermal conductivity of the suspended particle with respect to the liquid. From this observation, we propose a theoretical model of particle-particle interactions at the solid-liquid interfaces.
Essentially, the thermal conductivity mismatch leads to long-range deflection of the freezing interface near the particles. Then, the particles are repelled in a direction orthogonal to the local front interface, giving rise to an effective particle-particle interaction. This scenario bears striking similarities to the classical \textit{Cheerios effect}, which governs particle-particle interactions at liquid-air interfaces~\cite{vella2005cheerios}. We thus decided to name this effect the \textit{frozen Cheerios effect}. Obviously, the driving forces in the classical Cheerios case are different: the long-range deflection in the liquid-air case arises primarily from capillary forces, while the motion of the particles themselves is driven by buoyancy. Nevertheless, the essential ingredients for particle-particle interactions at an interface are similar in both cases, \textit{i.e.} long-range deformation of the interface and a repulsive force normal to it.

The paper is organized as follows. We first detail the experimental methods and sample preparation in Sec.\,\ref{Sec:Exp}, before discussing the experimentally observed particle-particle interaction in Sec.\,\ref{Sec:Results}. We then turn to the modeling of the frozen Cheerios effect in Sec.\,\ref{Sec:Cheerios}. We first derive the shape on the solid-liquid interface in Sec.~\ref{SubSec:Shape interface} and then the repelling speed and interaction length in Sec.~\ref{Sec:Interaction}. We end with concluding remarks in Sec.\,\ref{Sec:Conclusion}.

\section{Experimental methods}
\label{Sec:Exp}

\begin{figure}
  \centerline{\includegraphics[width=0.55\textwidth]{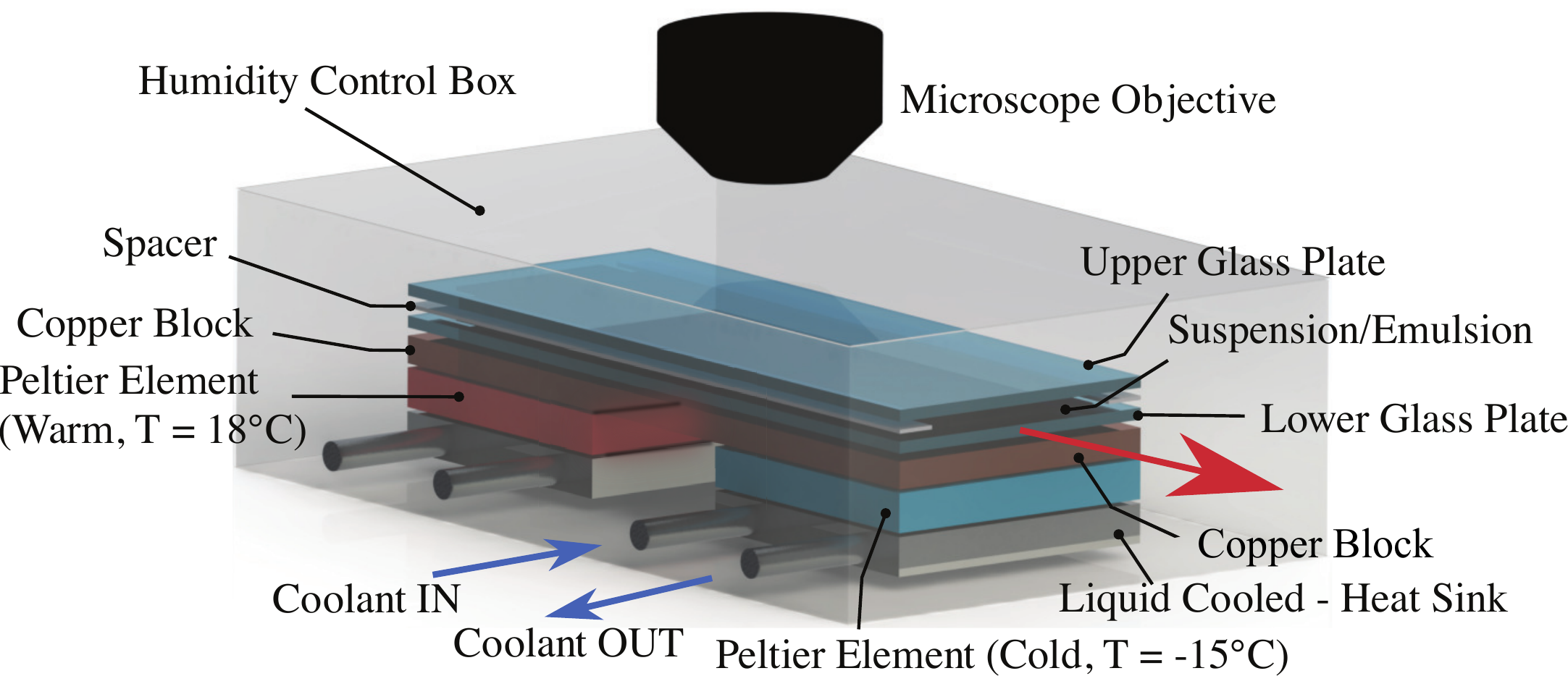}}
\caption{Schematic of the experimental set-up. The arrow indicates the direction in which the Hele-Shaw cell is pushed over the copper block. }
\label{fig:SI_1}
\end{figure}

\subsection{Unidirectional freezing set-up}

To study the interaction of particles at an advancing solidification front we perform unidirectional freezing experiments of dilute suspensions/emulsions. A schematic of the experimental set-up is shown in Fig.\,\ref{fig:SI_1}. Our experimental set-up uses a liquid cooled heat sink (Julabo) set to $\SI{-8.5}{\celsius}$ in combination with a Peltier element in order to reach a constant temperature at the cold side of $-15\pm0.2\,\SI{}{\celsius}$. A copper block, which temperature is monitored, is placed on top of this Peltier to distribute the heat equally. On the other (warmer) side we use an additional Peltier element underneath a second copper block to reach a constant temperature of $18\pm0.2\,\SI{}{\celsius}$. Both copper blocks are roughly $\SI{3}{\mm}$ apart resulting in a thermal gradient in the order of $G \approx \SI{1e4}{\kelvin \per \meter}$. We note that we did not precisely measure the temperature field within the cell and the estimation arises from the ratio between the typical temperature drop over the distance between the copper blocks. On top of these copper blocks a Hele-Shaw cell is placed (Ibidi, $\mu$-Slide I Luer with a thickness of $\SI{200}{\micro \meter}$), filled with our working liquids. The Peltier elements, the copper blocks and the Hele-Shaw cell sample are placed inside a humidity-controlled box to prevent fog and frost formation that would obscure the view and would induce local thermal conductivity inhomogeneities. Before starting the experiments we let the system equilibrate towards a constant and fixed thermal gradient $G$.  

We initiate the solidification on the colder end of the Hele-Shaw cell by inserting a small piece of ice, that acts as nucleation site. After the initialization, a solidification front forms that advances towards the warmer end, up to a position near the middle of the cell. This middle region, where the solid-liquid interface is located, is imaged with a Nikon D850 camera equipped with long working distance lens of $\SI{86}{\milli \meter}$. Once the front has reached a stable position, we set the Hele-Shaw cell in sliding motion at an imposed speed $V$, towards the colder end, using a high-precision linear actuator (Physik Instrumente, M-230.25). By doing so, the liquid in the vicinity of the solid-liquid interface immediately freezes such that the solid-liquid interfaces remains at a fixed position in the lab frame. Therefore, in the reference frame of the solid-liquid interface, the particles will approach the front at velocity $V$, while the thermal gradient remains fixed. We restrict ourselves to low freezing speeds (typically below $\SI{2}{\micro \meter \per \second}$) to prevent the emergence of so-called Mullins-Sekerka instabilities \cite{mullins1964stability} that may destabilize the front due to local perturbations of the melting temperature.  The lowest velocity that could be achieved is $V = \SI{0.1}{\micro \meter \per \second}$. The local variations of the front speed are of the order of 5\% of the imposed speed in such conditions. When increasing the imposed speed, the stability increases and the variations fall within 2\% for $V \approx \SI{1}{\micro \meter \per \second}$ and beyond. 

\subsection{Working fluids}

For our current investigations we consider oil-in-water emulsions and various (non)-aqueous suspensions. 
For the former, we use Milli-Q water for the bulk phase and silicone oil (Sigma-Aldrich, Germany) with a viscosity of $\SI{50}{cSt}$ as the dispersed phase. The emulsion is prepared using a co-flow microfluidic device.
To ensure stability of the emulsion a surfactant TWEEN-80 (Sigma-Aldrich, Germany) is added to the water prior to preparation with an initial surfactant concentration of $0.5\, \mathrm{vol}\%$.
Before starting the experiment, part of this sample is diluted by a factor 50 with Milli-Q water, resulting in a surfactant concentration of $0.01\, \mathrm{vol}\%$ in the sample. 
We have checked previously that the dilution factor has no influence in the particle-front interactions, which has been observed to occur at larger surfactant concentrations \cite{tyagi2022solute}, meaning that surfactants plays a minor role in the particle-front interactions in our experiments.  

Turning to the suspensions, we either use Milli-Q water or dimethyl sulfoxide (DMSO, Sigma Aldrich) for the bulk phase. The latter has a melting temperature of $T_m \approx \SI{16}{\celsius}$ and the fixed thermal gradient that we apply over this sample had to be adjusted for the solid-liquid interface to be in the middle of the cell ($0\pm0.2\,\SI{}{\celsius}$ on the cold side and $29\pm0.2\,\SI{}{\celsius}$ on the warm side).
The glass microspheres that are suspended in both bulk liquids are purchased from Whitehouse Scientific, UK and have a diameter variation between $38-45\, \SI{}{\micro \meter}$. 

For the remaining two aqueous suspensions we purchased polystyrene (PS) particles (Microbead Dynoseeds TS40) and polymethyl methacrylate (PMMA) particles (Cospheric microspheres, $38-45\, \SI{}{\micro \meter}$ in diameter). 
While preparing the suspensions, some particles showed clear attractive interactions by forming clusters when dispersed in water,  especially PS and PMMA particles. 
We found that by letting the particles be submerged in water for a couple of days and by frequently separating the particles through exposure to ultrasound, this initial clustering was not observed any more, allowing to get a nearly homogeneous initial distribution of particles in the Hele-Shaw cell prior to freezing.  The particle did not show any significant interactions in the liquid phase and were fully static (in the moving Hele-Shaw cell frame) prior making contact with the front.

\section{Particle-particle interactions at a freezing front}
\label{Sec:Results}

\begin{figure*}
  \centerline{\includegraphics[width=0.85\textwidth]{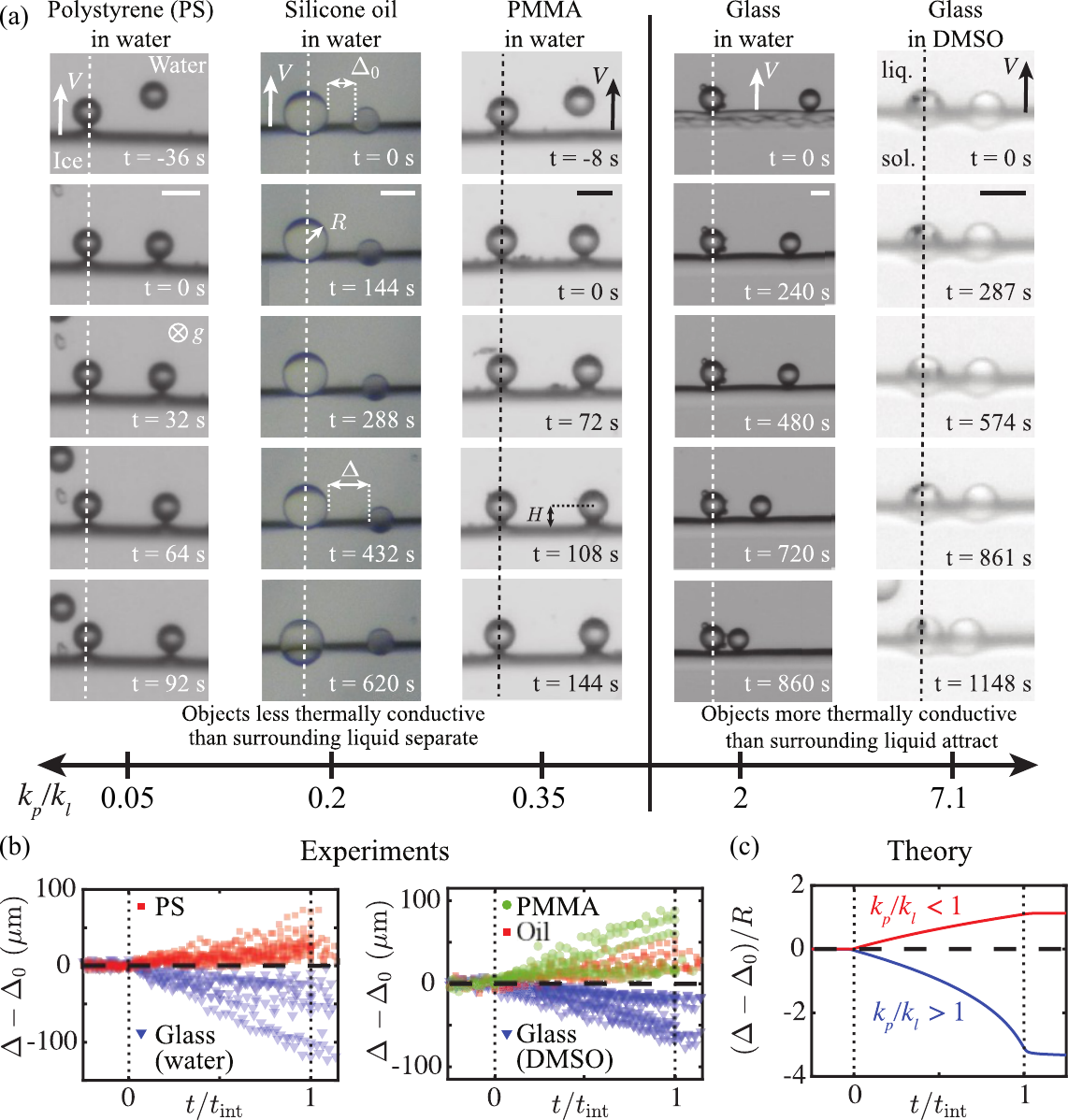}}
\caption{ Interactions between two particles at a moving solidification front. (a) Sequence of images capturing the relative dynamics of two (from left to right) polystyrene (PS) particles, silicone oil droplets, polymethyl methacrylate (PMMA) and glass particles in water, as well as two glass particles in dimethyl sulfoxide (DMSO), at an advancing solidification front. We put ourselves in the reference frame of the larger particles. The velocity at which the front propagates is slightly above the critical velocity $V_{\mathrm{crit}}$, that governs particle rejection or engulfment into the solidifying bulk.
The thermal conductivity of the particle and the surrounding liquid are denoted by $k_p$ and $k_l$.
The scale-bars indicate a length of $\SI{50}{\micro \meter}$. 
(b) Variation in shortest lateral distance between various particle pairs for each system, $\Delta(t) - \Delta_0$, as a function of normalised time. PS particles, oil droplets and PMMA particles are observed to repel whereas glass particles attract, in both water and DMSO. 
(c) Theoretical particle-particle interaction as a function of normalised time for two representative cases where $k_p/k_l = 2$ (blue) and $k_p/k_l = 0.2$ (red), corresponding to the thermal conductivity ratios of glass/water and PS/water respectively.}
\label{fig:1}
\end{figure*}

The sequences of images in Fig.\,\ref{fig:1}(a) display examples of pair interactions between particles at the freezing front. From left to right, the particles are respectively polystyrene (PS) particles, silicone oil droplets, polymethyl methacrylate (PMMA) and glass particles in water, as well as glass particles in dimethyl sulfoxide (DMSO). The snapshots in Fig.~\ref{fig:1}(a) are extracted from the Supp. Video 1-5. To highlight the relative distance $\Delta$ between the particles, the images are reframed such that the left particle is always at the same lateral and vertical position. In all cases, the time $t=0$ corresponds to the first time where both particles are in contact with the front. In our experiments, we could observe hundreds to thousands events of particle engulfment in ice. 
Among those, we proceed a selection of the specific cases where two particles are at the front at the same time, and where the initial distance between the particles is small enough to observe an effect, which in practice corresponds to typically 3 times the particle radius. 
Although the two particles might arrive at the front at $\textit{excalty}$ the same time, this does not necessary have to be the case. 
It might occur that particle A is being pushed by the front while particle B is still in the bulk. 
At some later time (typically in the order of seconds to minutes later), particle B might eventually make contact with the front, near A, after which the two particles start to interact.
For these situations we make sure that particles A and B were sufficiently isolated from any other particles, so that the interaction remains strictly binary.

The front speed was adjusted to be larger but close to the critical engulfment speed, typically of the order of $V\sim \SI{1}{\micro \meter \per \second}$, the exact value being system-dependent. First, we observe that for a duration of the order of a few minutes, the particles stay in contact with the moving front such that they get significantly displaced in the freezing direction, \textit{i.e.} traveling a distance of several times the particle radius. The particles start getting repelled by the front when they are near-contact, meaning for distances of the order of a few microns. The second observation is that the particles move in the transverse direction to the front in a nontrivial way. Indeed, in all cases where the thermal conductivity of the particles ($k_p$) is less than that of the surrounding liquid ($k_p < k_l$, left columns in Fig.~\ref{fig:1}(a)), the particles have a tendency to separate. On the other hand, for the case of more thermally conductive glass particles in water and in DMSO ($k_p > k_l$, right columns in Fig.~\ref{fig:1}(a)), the particles tend to attract up to contact. 
To be more quantitative, we plot in Fig.~\ref{fig:1}(b) the relative lateral displacement $\Delta(t) - \Delta_0$ between the particles, where $\Delta_0 = \Delta(t=0)$ is the initial particle-particle distance. The lateral motion of the particles stops either because they made contact, \textit{i.e.}, $\Delta = 0$ (for glass particles), or because at least one of the particles is being engulfed into the ice, or when the distance between the particles has become too large. We notice that in Fig.~\ref{fig:1}(b) we overlay all the observed pairwise interactions, which means 58 pairwise interactions, respectively 14, 8, 8, 12 and 16 for PS, Silicone Oil, PMMA, Glass in Water and Glass in DMSO. The time has been rescaled by the total interaction time for the sake of clarity. Clearly the attractive/repulsive nature of the interaction is systematic for a given particle-liquid system. 

However, we point out that in all cases the particles at the freezing front are subjected to significant fluctuations, as it can be appreciated in the Supplementary Movies 1-5. As a result, the relative displacements are also fluctuating, as evidenced in the spreading of the data in Fig.~\ref{fig:1}(b). This stochastic effect cannot be the result of Brownian-like thermal fluctuations given the size of the particle (typically $\SI{40}{\micro \meter}$). The motion of the individual particles seem to be very sensitive to local fluctuations of the environment during their interaction with the front, a feature that has already been observed in a similar experimental set-up \cite{dedovets2018five}. We believe that those fluctuations are not resulting from inhomogeneities in the thermal conductivity of the system, as the freezing front is quite stable. Understanding the nature of the underlying fluctuations would be an interesting challenge which however is beyond the scope of the present article. 

Finally,  we note that the observed effect cannot be due to bulk particle-particle interactions, as PS and PMMA particles tend to cluster in the bulk, while they separate when interacting at a freezing front. Similarly, glass particles usually separate in water owing to electrostatic repulsion, while they attracted here.
So in summary, we observe that the isolated particle-pairs either repel each other (PS, silicone oil and PMMA) or attract (glass in water and DMSO), where the attractive/repulsive nature of the interactions seems to be dictated by the mismatch in the thermal conductivities between the particle and the surrounding liquid. 

In the remainder of the article, we will focus on two aspects of those particle-particle interactions. First, we propose a possible mechanism of these particle-particle interactions, that is inspired by the \textit{Cheerios effect} for particle-particle interactions at liquid-air interface. In a second part, we discuss the interaction time/length of a single particle with the freezing front, that will dictate the magnitude of particle-particle interaction. 

\section{Frozen Cheerios model}
\label{Sec:Theory}

\subsection{Particle-particle interaction}
\label{Sec:Cheerios}

\begin{figure}
  \centerline{\includegraphics[width=0.8\textwidth]{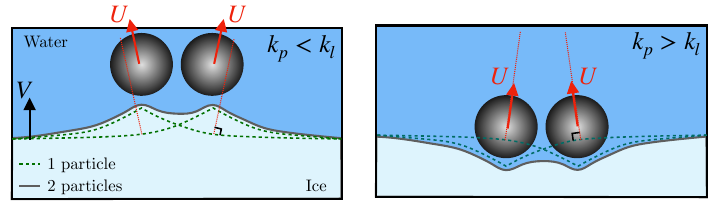}}
\caption{ Frozen Cheerios effect mechanism. Schematic representation of the particle-particle interaction at the moving solidification front in the case of particles less conductive than the surrounding liquid (left) and more conductive particles (right). The front deformation are exaggerated for pedagogical purposes. The green dashed lines indicate the shape of the solid-liquid interface for a single particle, while the solid gray line shows the total solid-liquid interface, assumed to be the sum of the two individual contributions. The red dotted lines show a line orthogonal to the perturbation of the interface provided by the other particle.  }
\label{fig:Cheerios}
\end{figure}

We suppose two particles at an identical distance from a freezing front moving with a velocity $V$ (Fig.\,\ref{fig:Cheerios}). We recall that PS/PMMA particles form clusters in the bulk owing to particle-particle interaction forces, while they repelled at the freezing front. Furthermore glass particles usually repelled each other due to electrostatic forces while they attract at the freezing point. For these reasons, we neglect bulk particle-particle interactions and only consider the interactions mediated by the freezing front. Due to intermolecular forces, each particle is repelled from the solid-liquid interface. The force balance between the repulsive force and the viscous drag sets the repelling speed $\mathbf{U}$, as it will be detailed in Sec.~\ref{Sec:Force}. Both of these effects involve length scales that are small compared to the sphere radius and the long-ranged deflection of the solid-liquid interface due to the thermal conductivity mismatch (justified in the following section). As a result, we do not expect the presence of a second particle to alter the particle-front interaction force, which, to leading order, is the same as the single particle-front dynamics. 

However, what the second particle does is to generate a long-range (of order of the particle radius $R$) deflection of the solid liquid interface. To a first approximation, we assume the deflection of the solid-liquid interface to be the sum of the individual front-particle deflections. With this approximation, each particle is getting repelled in the normal direction to its local environment, which is influenced by the other particle. Then, the repelling velocity of the one particle is oriented in the normal direction to the long-range deformation induced by the other particle (see red dotted lines in Fig.~\ref{fig:Cheerios}).
This is the cause of the effective particle-particle interactions. 
More precisely, the tangential velocity can be express by 
\begin{equation}
\label{eq:Cheerios}
U_r =  -U_z \frac{\partial z}{ \partial r},
\end{equation}
where $\frac{\partial z}{ \partial r}$ is the local slope of the interface at the center of the second particle,  imposed by a neighbouring particle.
The sign of the slope changes from negative to positive for less ($k_p < k_l$) to more ($k_p > k_l$) thermally conductive particles, respectively,  and so does the repulsive/attractive nature of the particle-particle interaction. 
The lateral displacement of the particles can then be found by integrating the lateral velocity Eq.\,(\ref{eq:Cheerios}) as 
\begin{equation}
\Delta(t) - \Delta_0 = \int_0^t U_r \, \mathrm{d}t,
\label{eq:latdisplacement}
\end{equation}
where $\Delta(t)$ is the lateral particle-particle distance and $\Delta_0 = \Delta(t =0)$ the initial distance.\\

In the following sections we will derive in detail both components of Eq.\eqref{eq:Cheerios} that are required to understand the particle-particle interaction. We first focus on the shape of the interface away from the particle in Sec.\,\ref{SubSec:Shape interface}) to obtain the interface slope $\frac{\partial z}{ \partial r}$. 
An asymptotic expression will be provided for the far-field front deflection.
Then the particle repelling speed normal $U_z$ to the planar front will be derived in Sec.\,\ref{Sec:Force}. 
Combining Eqs. (1) and (2) together with the far-field asymptotic front deflection, we obtain a theoretical prediction of the lateral particle displacement $\Delta(t) - \Delta_0$ plotted in Fig.\,\ref{fig:1}(c), which agrees qualitatively with the average behaviour of the particles observed experimentally (Fig.\,\ref{fig:1}(b)).  

\subsection{Shape of the solid-liquid interface for an isolated particle}
\label{SubSec:Shape interface}

\begin{figure}[t!]
  \centerline{\includegraphics[width=0.4\columnwidth]{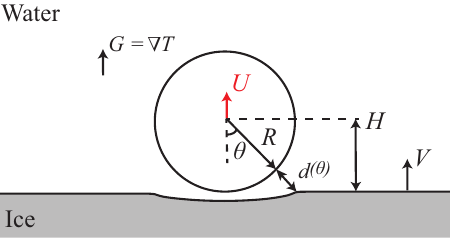}}
\caption{Sketch of an isolated particle near a freezing front. We consider an axisymmetric, spherical coordinate system centered on the particle. }
\label{fig:SI_2}
\end{figure}

We now determine the shape of the freezing interface, that is needed to obtain the local slope in the frozen Cheerios model. To that end, we closely follow the previous works dedicated to the interaction between an isolated particle with a solidification front~\cite{rempel1999interaction,rempel2001particle,park2006encapsulation}. We recall that the key assumption is that the deflection of the solid-liquid interface with two particles can be approximated as the sum of the two one-particle deflections (see Fig.~\ref{fig:Cheerios}). Therefore, we consider here only the shape of the solid-liquid interface for an isolated particle, as schemed in Fig.~\ref{fig:SI_2}.

Under standard conditions the solid-liquid interface temperature $T_i(\mathbf{x}) $ is given by the melting temperature of water $T_m$. In close proximity to the particle, however, the equilibrium condition changes due to local variations in pressure that modify the melting temperature. The interface temperature is then given by the Gibbs-Thomson relation \cite{perez2005gibbs}
\begin{equation}
T_i(\mathbf{x}) = T_m - \frac{T_m}{\rho_s \mathcal{L}} \, \Delta p(\mathbf{x}), \quad  \text{with} \quad \Delta p(\mathbf{x}) = \sigma_{\mathrm{sl}} \mathcal{K}(\mathbf{x}) + \Pi(\mathbf{x}),
\label{Eq:EquilibiurmEq}
\end{equation}
where $\rho_s$ is the mass density of ice, $\mathcal{L}$ the latent heat, $\sigma_{\mathrm{sl}}$ is the surface energy of the solidification front, $\mathcal{K}$ its local curvature and $\Pi$ the disjoining pressure. Particles engulfed in ice are known to be surrounded by a premelted liquid layer coming from intermolecular interactions~\cite{wettlaufer2006premelting}. We suppose that these interactions are the main cause of the particle repulsion by the moving front, and assume the presence of a van-der-Waals disjoining pressure, as $\Pi(\mathbf{x}) =  A/(6\pi d^3(\mathbf{x}))$, where $A$ is the Hamaker constant and $d(\mathbf{x})$ the local particle-front distance.  At this point, we point out that the present model treats the particle-front intermolecular interaction in a fairly simple way, while more complex models have already been proposed~\cite{kao2009particle}, modeling more closely the details of the interactions. We wish here rather to focus on the particle-particle interactions, and the Hamaker constant will be fitted in order to match the experimentally observed critical engulfment velocity (see Sec.\,\ref{Sec:Interaction}).

Then, the heat equation allows to get a second condition that fully defines the solid-liquid interface. We assume that thermal diffusion dominates over convection \footnote{the Peclet number is $\textit{Pe} = UR/\kappa_T \approx 10^{-4}$, with typical values for velocity $U = V \sim 10^{-6} \, \SI{}{\meter \per \second}$, particle size $R \sim 10^{-5} \, \SI{}{\meter}$ and thermal diffusivity $\kappa_T \sim 10^{-7} \, \SI{}{\meter \squared \per \second}$.}, and the temperature field to be steady. Hence, the temperature field satisfies Laplace's equation $\nabla^2 T = 0$ in both phases. An external far-field thermal gradient $\mathbf{G} = G\mathbf{e}_z$ is applied, and a spherical particle of radius $R$ is placed at a distance $H$ from the reference undeformed front, where its center of mass is at $\mathbf{x}_0$. The thermal conductivity of the particle and the surrounding liquid are denoted by $k_p$ and $k_l$. For simplicity, the ice is assumed to have the same conductivity as water, \textit{i.e.}, $k_i = k_l$, whereas in reality $k_i/k_l \approx 4$. Full numerical simulations of the heat equation with $k_i/k_l = 4$ have shown no qualitative difference of the interface with the current model \cite{van2024deforming}. This assumption allows us to use analytic solutions of the heat equation~\cite{shangguan1992analytical} and the solid-liquid interface corresponds to the isotherm that matches the melting temperature in the far field, which is given by
\begin{equation}
T_i(\mathbf{x}) = T_m + GH + \mathbf{G} \cdot (\mathbf{x}-\mathbf{x}_0)  + k_eR^3  \, \frac{ \mathbf{G} \cdot (\mathbf{x}-\mathbf{x}_0)  }{\vert \mathbf{x} - \mathbf{x}_0 \vert^3} ,
\label{Eq:ThermOuter}
\end{equation}
with $k_e = (k_l-k_p)/(2k_l+k_p)$, with $-1<k_e<1/2$, is the dimensionless parameter that quantifies the mismatch in the thermal conductivities. Importantly, when the particle and liquid have different thermal conductivities, $k_e \neq 0$, the solid-liquid interface is shifted because of the heat flux boundary condition at the particle surface, leading to the extra term of dipolar symmetry in \eqref{Eq:ThermOuter}. 

Combining equations \eqref{Eq:EquilibiurmEq} and \eqref{Eq:ThermOuter} yields an expression for the shape of the solidification front that reads
\begin{equation}
\frac{T_m}{\rho_s \mathcal{L}} \, \left[ \sigma_{\mathrm{sl}} \mathcal{K}(\mathbf{x}) + \Pi(\mathbf{x}) \right] + GH + \mathbf{G} \cdot (\mathbf{x}-\mathbf{x}_0)  + k_eR^3  \, \frac{ \mathbf{G} \cdot (\mathbf{x}-\mathbf{x}_0)  }{\vert \mathbf{x} - \mathbf{x}_0 \vert^3} = 0.
\label{Eq:SI_Interface}
\end{equation}
We introduce the spherical coordinate system centered on the particle, see Fig.\,\ref{fig:SI_2}, such that the shape of the solid-liquid interface is determined via the local particle-front distance $d(\theta)$ as
\begin{equation}
\label{eq:front-shape}
L_{\sigma} \mathcal{K} R + \frac{L_{\Pi}^4}{d^3(\theta)} = (R + d(\theta)) \cos \theta \left[ 1 + k_e \left( \frac{R}{R+d(\theta)} \right)^3 \right] - H,
\end{equation}
where we introduce the length scales $L_\sigma$ and $L_\Pi$ associated with capillary effect and the disjoining pressure, as done in Refs.~\cite{rempel1999interaction,rempel2001particle,park2006encapsulation}:
\begin{align}
L_{\sigma} & = \frac{T_m \sigma_{\mathrm{sl}}}{\rho_s \mathcal{L}G R} , & L_{\mathrm{\Pi}} &=  \left(\frac{ A T_m}{6 \pi \rho_s \mathcal{L} G} \right)^{1/4}.
\end{align}
The typical film thickness near the basis of the particle $d^*$ can be found balancing the disjoining pressure with the Laplace pressure as $A/(6\pi d^{*3})\sim \sigma_{\mathrm{sl}}/R$, which is also related to the aforementioned lengths scales via $d^* = \left(L_{\Pi}^4/L_{\sigma} \right)^{1/3}= R[A/(6\pi R^2\sigma_{\mathrm{sl}})]^{1/3}$. 
Substituting typical values results in a thickness in the order of tens to hundreds of nanometers, depending on the precise value of the Hamaker constant (see Sec.~\ref{SingleParticleDisp}).
Defining the solid-liquid interface with the front-particle distance $d$, the curvature of the former follows as
\begin{equation}
\mathcal{K} = \left[ \frac{-2(R+d)^2 + (R+d)d^{\prime \prime} - 3d^{\prime 2}}{((R+d)^2 + d^{\prime 2})^{3/2}} + \frac{\cos \theta }{\sin \theta}\frac{d^{\prime}}{(R+d)((R+d)^2+d^{\prime 2})^{1/2}} \right],
\label{Eq:kappa}
\end{equation}
with primes denoting derivatives with respect to $\theta$. 
Equations~(\ref{eq:front-shape}) and (\ref{Eq:kappa}) are made dimensionless and solved numerically for any particle-front distance $H$ (see Appendix~\ref{app_sec:num} for additional details) \footnote{The numerical code can be found at \url{https://github.com/vincent-bertin/frozen_cheerios}}. 

\begin{figure}
  \centerline{\includegraphics[width=0.85\textwidth]{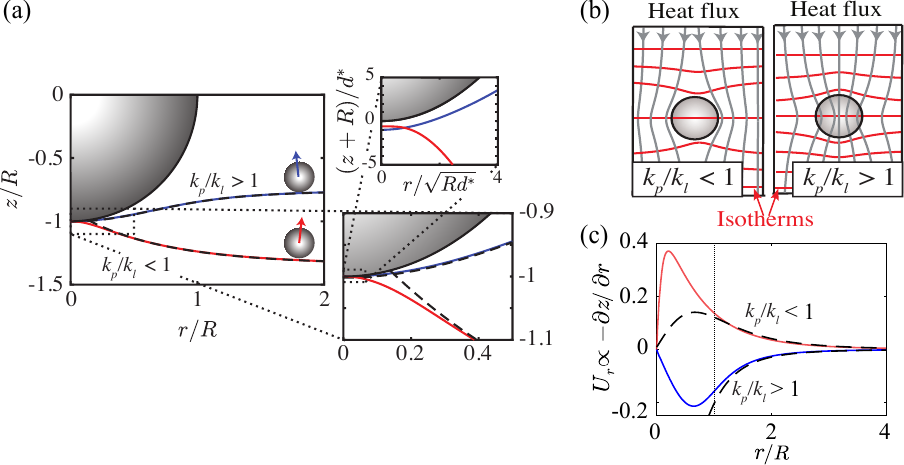}}
  \caption{Solid-liquid interface shape.
  (a) Numerical solution of the solid-liquid interface for $H = R(1 + k_e)$, with a conductivity ratio of $k_p/k_l = 2$, \textit{i.e.}, $k_e = -0.25$, (blue) and $k_p/k_l = 0.2$, \textit{i.e.}, $k_e = 0.36$ (red). A neighbouring particle might interact with the inclined interface, causing them to attract or repel, as shown with the schematic. A thin liquid film of typical thickness $d^*$ is separating the particle from the solid-liquid interface, as seen in the top right panel, which is a rescaled zoom-in of the first zoom-in. The dashed black lines correspond to the asymptotic limit away from the particle, \textit{i.e.}, Eq.\,(\ref{eq:limit}), that only deviate from the full solution at its base. 
(b) Qualitative interpretation of the front deflection due to conductivity mismatch. Schematic of the heat flux streamlines (gray) and isotherms (red) when a particle is placed in an uniform external temperature gradient. In the left (resp. right) panel, the particle is less (resp. more) conductive than the surrounding liquid. 
(c) Negative slope of the interface versus $r/R$ that dictates the attractive/repulsive nature of the particle-particle interaction (see Eq.(\ref{eq:Cheerios})). The dashed lines correspond to the solutions of the linearly expanded asymptotic limit, \textit{i.e.}, Eq.\,(\ref{eq:long-range_result}). The dotted line indicates $r/R = 1$, the distance at which equally sized particles will touch.}
\label{fig:3}
\end{figure}

The shape of the solid-liquid interface is shown in Fig.~\ref{fig:3}(a) for both more ($k_p/k_l = 2$, see blue lines) and less ($k_p/k_l = 0.2$, see red lines) conductive particles when the front-particle distance $H = R (1+k_e)$. The latter corresponds to the typical distance at which particle and ice nearly touch. A thin liquid film, of typical thickness $d^*$, is separating the particle from the interface. Such deflections of the solidification front have already been observed experimentally \cite{tyagi2020objects,van2024deforming}, and at low freezing speeds the advancing front deflects as depicted in Fig.\,\ref{fig:3}(a).

\subsection*{Qualitative interpretation of the front deflection due to conductivity mismatch }

We provide here a qualitative explanation about the effects of the thermal conductivity mismatch on the solid-liquid interface. To that end, we simplify the discussion and consider a system with a uniform external temperature gradient. Without any particle, the heat flux streamlines are straight and in the direction of the temperature gradient, via Fourier’s law. When a particle of different thermal conductivity is placed in the system, the heat flux streamlines will bend in the vicinity of the particle. If the particle is less conductive, the heat flux will be deflected away from the particle as the resistance is locally higher near the particle (see left in Fig.~\ref{fig:3}(b)). The isotherms being orthogonal to the heat flux, they bend towards the particle. The opposite occurs for more conductive particles, such that the isotherms bend away from the particles. The solid-liquid interface being an isotherm at the melting temperature, we observe similar front deflections. 

\subsection*{Shape of the solid-liquid interface at large distance}

The frozen Cheerios model requires to know the shape of the solid-liquid interface at distances larger than the particle radius. As the particle-front interactions are only significant at short distances (near the base), they do not affect the long-distance solid-liquid interface. Furthermore, the far-field deformation is entirely dictated by the thermal conductivity mismatch, via the dipolar term in \eqref{Eq:ThermOuter}. Indeed, far from the base of the sphere, the short-range disjoining pressure, governing the solution of the inner region, can be neglected and the interface position follows 
\begin{equation}
\label{eq:long-range}
\frac{T_m\sigma_{\mathrm{sl}}}{\rho_s \mathcal{L}} \mathcal{K}(\mathbf{x}) + GH + \mathbf{G} \cdot (\mathbf{x}-\mathbf{x}_0)  + k_eR^3  \, \frac{ \mathbf{G} \cdot (\mathbf{x}-\mathbf{x}_0)  }{\vert \mathbf{x} - \mathbf{x}_0 \vert^3} = 0.
\end{equation}
Eq.\,\eqref{eq:long-range} compares the solid-liquid Laplace pressure to the thermal gradient, which has a similar form as the hydrostatic pressure. Therefore, a length scale arises comparing these two pressures that is analogous to the capillary length in a gravito-capillary system, as 
\begin{equation}
\sqrt{\frac{T_m\sigma_{\mathrm{sl}}}{\rho_s \mathcal{L} G}} = \sqrt{L_\sigma R}.
\end{equation}
For water-ice, with a thermal gradient $G = \SI{1e4}{\kelvin \per \meter}$ and surface energy $\sigma_{\mathrm{sl}} = \SI{0.03}{\joule \per \meter \squared}$ \cite{rempel1999interaction}, the thermal capillary length is of the order of $\SI{1}{\micro \meter}$, and is much smaller than the particle radii in our experiments. Therefore, capillary effects are not predominant here at large distances, as the typical curvature of the interface is $1/R \ll 1/\sqrt{L_\sigma R}$. As a result, the capillary term can be neglected as well at large distance. The long-range deformation of the interface is more conveniently expressed with cylindrical coordinates, where $r$ is the radial position. The interface shape is denoted $z(r)$, which is defined with respect to the center of the particle, \textit{i.e.} $z = (\mathbf{x}-\mathbf{x}_0) \cdot \hat{z}$. The Eq.\,\eqref{eq:long-range} in cylindrical coordinates gives 
\begin{equation}
\label{eq:limit}
H + z + k_e R^3 \frac{z}{(z^2+r^2)^{3/2}}=0.
\end{equation}
Lastly, performing an linear expansion around the flat interface as $z = -H + \delta z$, with $\delta z \ll H$, we find 
\begin{equation}
\label{eq:long-range_result}
z \simeq - H +k_e R^3 H/(H^2+r^2)^{3/2}.
\end{equation}
These simplified expressions match very well the full numerical solutions of the Gibbs-Thomson Eq.\,\eqref{Eq:SI_Interface} at larger distances, as shown in Fig.~\ref{fig:3}(a)\&(c) (dashed lines). The sign of the interface slope changes with $k_e$, corresponding to the transition from particles less to more conductive than the surrounding liquid. 
We indeed find that the typical range of the solid-liquid interface shape is given by the particle radius. The interface slope decays algebraically with the radial distance as $\partial z/ \partial r \simeq k_e R^3 H/r^4$, which is a fairly fast decay. That justifies why particle-particle interactions are only effective when particles are sufficiently close (typically 3 particle radius in the experiment). 

We notice that the deflection of the solid-liquid interface in the outer region is very different from the one derived by Park et al (see Eq. (2.24) in Ref.~\cite{park2006encapsulation}). The reason being that Park et al. performed an expansion around a flat solid-liquid interface and scaled the film thickness with $d^*$. In contrast, we found that the typical scale of the deflection of the solid-liquid interface due to the thermal conductivity mismatch is $k_e R$, which largely exceeds $d^*$ in practice. The expression derived here accurately describes the full numerical resolution of the Gibbs-Thomson relation (see Fig.\,\ref{fig:3}(a)) which supports that Eq.\,\eqref{eq:long-range_result} indeed correctly accounts for the far-field deflection of the solid-liquid interface. 

For the sake of completeness, we comment on the specific case of particles having the same thermal conductivity of water, leading to $k_e \approx 0$. This limit could be relevant for biological objects (e.g. vesicles, cells) that are mainly made of water. In this situation, the Eq. \eqref{eq:long-range} reduces to $\frac{T_m\sigma_{\mathrm{sl}}}{\rho_s \mathcal{L}} \mathcal{K}(\mathbf{x}) + GH + \mathbf{G} \cdot (\mathbf{x}-\mathbf{x}_0) = 0$, corresponding to a balance between the Laplace pressure and a thermal term that is equivalent to an hydrostatic pressure.  Therefore, the shape of the interface is completely equivalent to the one in classical Cheerios effect in the limit $k_e = 0$, where the long-range deformation would be given by $z(r) \propto -d^* K_0\left(\frac{r}{\sqrt{L_\sigma R}}\right) \underset{r\gg \sqrt{L_\sigma R}}{\sim}  -d^* \frac{\sqrt{L_\sigma R}}{r},$ where $K_0$ is the modified Bessel function of the first kind. We do not provide here the exact prefactor that depends on the details of the profile, but the latter scales with the typical film thickness scale $d^*$. However, the front deflection would be negative as the intermolecular interactions are repulsive. Therefore, the specific case of $k_e =0$ would lead to attractive Cheerios interactions. However, we notice that the length scales associated with the typical slope deflection ($d^*$ and $\sqrt{L_\sigma R}$) are in the micron (or hundreds of nanometers) range, so we also expect Cheerios-like interactions to be small for particles with $R\approx 10\,\mu\text{m}$.

We have characterized the shape of solid-liquid interface deflection due to the presence of an isolated particle. Importantly, we have shown that the long-range deflection of the solid-liquid interface shape does not depend on the specific particle-front molecular interactions and is entirely set by the thermal conductivity mismatch. In particular, the front bends towards (resp. away from) the particle in the case of particles less (resp. more) conductive than the surrounding liquid (see Fig.~\ref{fig:3}(b)). Hence, within the frozen Cheerios scenario of Sec.~\ref{Sec:Cheerios}, particles less conductive than the surrounding liquid repel each other, while the opposite occurs for more conductive particles, as observed experimentally (see Fig.~\ref{fig:1}).
However, to make a theoretical prediction of the particle-particle interactions, we additionally need to know the extend to which particles interact with the moving front, which will be the main focal point of the following section.

\section{Interaction length: how much are the particles repelled?}
\label{Sec:Interaction}

The particle-particle interactions in the frozen Cheerios effect arises because of the repulsion of particles by the moving front prior engulfment. The goal of this section is to predict the interaction length between the particles and solidification front, or saying differently: how much are the particles repelled before getting engulfed. The interaction length will eventually dictate how much particles can move in the lateral direction along the deflected interface, discussed above.

\begin{figure}
  \centerline{\includegraphics[width=0.6\textwidth]{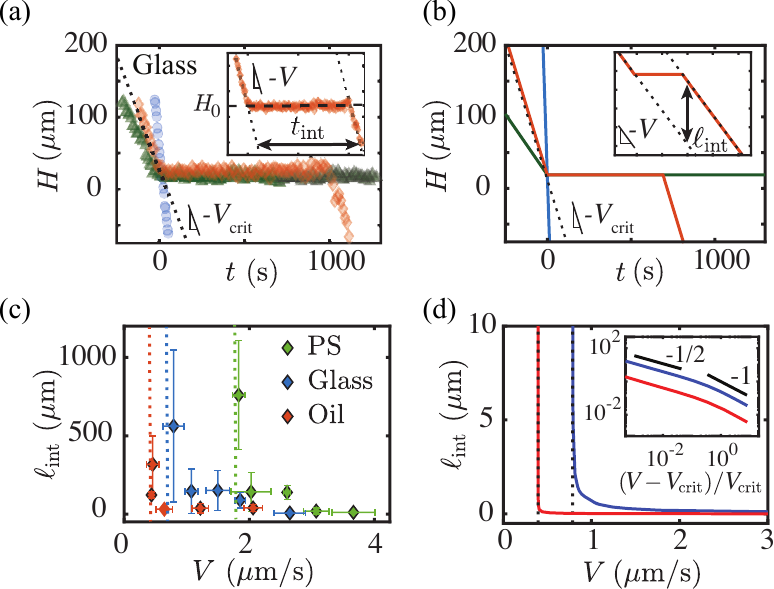}}
\caption{Freezing of a single particle. Front-particle distance $H(t)$ during the freezing of an isolated glass particle in water for experiments (a) and theory (b). Depending on the front velocity, three typical scenario are observed: full rejection (\protect \tikztriangle[triangle]{mygreen,fill=mygreen,opacity=0.4}) at low freezing rate, no displacement (\protect \tikzcircle[myblue2, fill=myblue2, opacity=0.4]{2.75pt}) for fast freezing, and partial rejection (\protect \tikzsymbol[diamond]{myorange,fill=myorange,opacity=0.4}) at intermediate speed (see Supplementary Movie 6). The insets highlight the definition of the particle-front interaction time $t_{\mathrm{int}}$ and length $\ell_\mathrm{int}$. 
(c) Experimental particle-front interaction length as a function of advancing velocity $V$ for single oil drops (\protect \tikzsymbol[diamond]{black,fill=myred}), glass (\protect \tikzsymbol[diamond]{black,fill=myblue}) and PS (\protect \tikzsymbol[diamond]{black,fill=mygreen2}) particles in water. The interaction length rapidly increases as $V$ approaches $V_{\mathrm{crit}}$ (dotted lines), where $V_{\mathrm{crit,oil}} \approx \SI{0.4}{\micro \meter \per \second}$, $V_{\mathrm{crit,glass}} \approx \SI{0.8}{\micro \meter \per \second}$ and $V_{\mathrm{crit,PS}} \approx \SI{1.8}{\micro \meter \per \second}$. 
Only the Hamaker constant $A$ is fitted to match the critical velocity $V_{\mathrm{crit}}$ in the theoretical model. 
We find $A_{\mathrm{oil}} \sim \SI{1e-20}{\joule}$ and $A_{\mathrm{glass}} \sim \SI{1e-18}{\joule}$ .
(d) Theoretical particle-front interaction length as a function of the advancing velocity $V$ for representative cases where $k_p/k_l = 2$ (blue) and $k_p/k_l = 0.2$ (red).
The inset on a log-log scale in highlights the critical behaviour of $\ell_{\mathrm{int}} \propto (V-V_\mathrm{crit})^{-1/2}$.}
\label{fig:2}
\end{figure}

\subsection{Experimental results}
\label{subsec:Experiment_ell_int}
We start this section with the experimental observations. The Supplementary Video 6 displays the interaction between glass particles and a solidification front moving at three different speeds, corresponding to the three asymptotic limits: slow freezing $V< V_\mathrm{crit}$, freezing near the critical speed $V \gtrsim V_\mathrm{crit}$ and fast freezing $V\gg V_\mathrm{crit}$. We introduced the critical engulfment speed $V_\mathrm{crit}$. Clearly, the interactions between a single particle and a planar solidification front depend strongly on the front velocity~\cite{rempel1999interaction,rempel2001particle,park2006encapsulation}. To illustrate this further, we show in Fig.~\ref{fig:2}(a) the sphere-front distance as a function of time for three cases in those respective limits. If the front velocity is smaller than a critical velocity $V_\mathrm{crit}$ (left case in Supp. Video 6), the particles are indefinitely repelled, and the front distance reaches a stationary value (green triangles in Fig.~\ref{fig:2}(a)). Oppositely, in the fast freezing limit, the front rapidly sweeps over the particle without any displacement, such that the particle keeps the same speed, \textit{i.e.} $-V$ in the reference frame where the freezing front is immobile (blue circles in Fig.~\ref{fig:2}(a)). 

The most interesting regime, meaning that would give rise to a strong frozen Cheerios effect, is when the front speed is near the critical engulfment speed. In that case, the particle is repelled by the front for a given time $t_\mathrm{int}$ before getting engulfed (see red diamonds in Fig.~\ref{fig:2}(a)), ultimately yielding to a particle displacement $\ell_\mathrm{int}$ (called later interaction length). We have performed multiple experiments by varying systematically the front speed to investigate how the interaction length varies with front speed, as shown in Fig.~\ref{fig:2}(c). The critical speed (see vertical dotted lines) depends on the type of particles. In all cases, the interaction length diverges as the front velocity approaches the critical velocity (Fig.~\ref{fig:2}(c)) in a very similar fashion for oil droplets, glass and PS particles in water. 
As we approach this critical point the scatter in the experimental data becomes more significant and is reflected in the size of the error bars in Fig.\,\ref{fig:2}\,(c).
We suspect that the increase in scatter arises due to the earlier mentioned intrinsic local fluctuations when the particle interacts with the front (see Sec.\,\ref{Sec:Results}). 
The longer the particle stays at the front, the more susceptible it is to these fluctuations.
In what remains, we will try to predict theoretically this interaction length. 
Eventually this will allow us to make theoretical predictions on the particle-particle interactions and compare those to our experimental observations, as mentioned in Sec.\,\ref{Sec:Cheerios}.

\subsection{Forces balance on the particle}
\label{Sec:Force}

To obtain the interaction length, the first challenge then is to predict the repelling speed $\mathbf{U}$ of a particle near a solidification front. As already discussed in various previous works, the repelling speed may be obtained by considering the force balance on the particle~\cite{shangguan1992analytical,rempel1999interaction,rempel2001particle, park2006encapsulation,garvin2007multiscale,tao2016steady}. First, the repelling force $\mathbf{F}_{\Pi}$ can be found by integrating the disjoining pressure on the particle surface. Then, the moving particle experiences a viscous friction force $\mathbf{F}_\mathrm{vis}$ opposing the particle motion, hence effectively attracting the particle towards the ice and promoting particle engulfment. The lubrication approximation is commonly used to compute the friction force in that context. The particle repelling speed is then determined from the force balance $\mathbf{F}_{\Pi}+\mathbf{F}_\mathrm{vis}=\mathbf{0}$. 

Both the viscous drag and the intermolecular forces are mainly set by the intimate region near the basis of the particle at length scale $\sim d^*$. Hence, we suppose that the second particle at distance $\approx R \gg d^*$ does not influence the leading order force balance and focus here again on isolated particle-front interaction. We shall point out some additional simplification assumptions done here (see Appendix~\ref{app:limitations} for more details). In the experiments, the particles are not neutrally buoyant and are in contact with the side wall of the Hele-Shaw cell. For the sake of simplicity, we ignore those side effects and assume an infinite system in the plane of solidification. Additionally, in the case of droplets, thermocapillary effects act on droplets placed in a thermal gradient, leading to a migration towards the warm side~\cite{young1959motion,subramanian1983thermocapillary,meijer2023thin}. We again neglect those possible effects. Lastly, the effects of water expansion during freezing will also be neglected. 

\subsection*{Friction force: viscous lubrication model}

We use again the axisymmetric spherical coordinate system centered on the particle (see Fig.\,\ref{fig:SI_2}), where $\theta$ is the latitude angle which is defined from the basis of the particle. Although the solid-liquid boundary is translated at a speed $V$, it does not contribute to the friction force, as the liquid particles near the solid-liquid boundary are not moving but freeze. Assuming a no-slip condition, the boundary condition for the velocity field at the solid-liquid surface is $\mathbf{v}=\mathbf{0}$~\cite{davis2001theory}. On the other hand, the boundary condition for the velocity field at the particle surface is $ \mathbf{v}=\mathbf{U}$, where $\mathbf{U}$ is the particle velocity. As seen in Sec.~\ref{SubSec:Shape interface}, the solid-liquid interface is not flat, but curved near the particle and characterized by the local front-particle distance $d(\theta)$ (see Fig.~\ref{fig:SI_2}). The friction force experienced by the particle depends on the precise shape of the solid-liquid interface.

Since the intermolecular interactions are effective on a length scale small compared to the particle radius, the particle repelling velocity is significant only when the particle is very close to the solid-liquid interface. As discussed in Sec.~\ref{SubSec:Shape interface}, the relevant scale for the film thickness at the basis of the particle is $d^*\sim R[A/(6\pi R^2\sigma_{\mathrm{sl}})]^{1/3} \ll R$, which is much smaller than the particle radius. In this situation, the viscous friction force is largely enhanced as compared to the classical Stokes drag and can be computed using the viscous lubrication approximation~\cite{batchelor1967introduction}. In this approximation, the viscous pressure only depends on the angle and is uniform in the normal direction to the interface. The viscous pressure $p_\mathrm{vis}$ follows from the thin-film equation
\begin{equation}
\label{eq:lubrication}
\mathbf{U}\cdot\mathbf{n} = \nabla \cdot \left( \frac{d^3}{12\mu} \nabla p_\mathrm{vis}\right),
\end{equation}
where $\mathbf{n}$ is the normal vector to the interface at a given angle. Using the spherical coordinate system, this yields 
\begin{equation}
 U  \cos \theta = \frac{1}{R^2\sin\theta} \frac{\partial}{\partial \theta} \left( \sin\theta \frac{d^3(\theta)}{12\mu} \frac{\partial p_\mathrm{vis}(\theta)}{\partial \theta} \right),
\end{equation}
where we recall that $d$ and $p_\mathrm{vis}$ only depend on the latitude angle. The latter equation can be integrated from the pole $\theta = 0$ (the basis of the particle) to an arbitrary latitude $\theta$ to give 
\begin{equation}
\label{eq:pressure-gradient}
\frac{\partial p_\mathrm{vis}(\theta)}{\partial \theta}  = 6 \mu U R^2  \frac{\sin\theta}{d^3(\theta)}. 
\end{equation}
The viscous pressure vanishes at large distances, \textit{i.e.} in the $\theta \rightarrow \pi/2$ limit, so that the Eq.\,\eqref{eq:pressure-gradient} can further be integrated over $\theta$ to obtain the viscous pressure field as 
\begin{equation}
p_\mathrm{vis}(\theta) = 6 \mu U R^2 \int_\theta^{\pi/2} \frac{\sin\hat{\theta}}{d^3(\hat{\theta})} \mathrm{d}\hat{\theta},
\end{equation}
where the integration constant (reference pressure) is set to zero. Finally, the friction force in the viscous lubrication theory is given by the integral of the viscous pressure on the surface, as
\begin{equation}
F_\mathrm{vis}(\theta) = 2\pi R^2 \int_0^{\pi/2} p_\mathrm{vis}(\theta)  \cos\theta\sin\theta \, \mathrm{d}\theta = 6\pi\mu U R^4 \int_0^{\pi/2} \mathrm{d}\theta \sin 2 \theta  \int_\theta^{\pi/2}  \frac{\sin\hat{\theta}}{d^3(\hat{\theta})} \, \mathrm{d}\hat{\theta},
\label{Eq:SI_Fvis}
\end{equation}
where we recover the expression as in Ref. \cite{park2006encapsulation}. 

\subsection*{Repelling force}

The repelling force, arising from the intermolecular interaction, is similarly obtained by integrating the disjoining pressure $\Pi(\theta) = A/(6\pi d^3(\theta))$ over the surface of the particle, as
\begin{equation}
F_{\Pi}(\theta) = 2\pi R^2 \int_0^{\pi/2} \Pi(\theta)  \cos\theta\sin\theta \, \mathrm{d}\theta = \frac{AR^2}{6} \int_0^{\pi/2}\frac{\sin 2 \theta }{d^3(\theta)}\mathrm{d}\theta.
\label{Eq:SI_Fpi}
\end{equation}

\subsection{Single particle displacement}
\label{SingleParticleDisp}
\begin{figure}
  \centerline{\includegraphics[width=0.65\columnwidth]{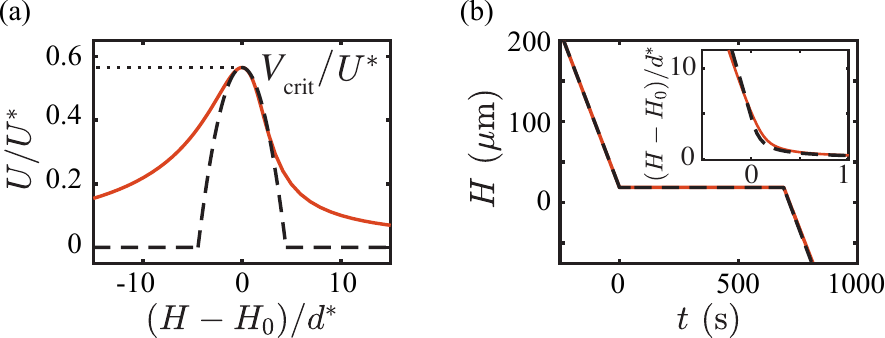}}
  \caption{ Particle repelling velocity. (a) Theoretical dimensionless particle velocity $U/U^*$ as a function of the rescaled particle-front distance $(H-H_0)/d^*$. The critical velocity corresponds to the maximum of the particle velocity. The dashed lines show the approximated velocity used in the simplified theoretical model. 
(b) Numerical (solid line) and analytical (dashed line) front-particle distance $H(t)$ as function of time.  The inset shows a rescaled zoom-in.}
\label{fig:4}
\end{figure}

Injecting the shape of the solid-liquid interface from Eq.~\eqref{eq:front-shape} into Eqs.\eqref{Eq:SI_Fvis} and \eqref{Eq:SI_Fpi}, we can compute both the repelling and the friction force and solve for the particle speed. At the scaling level, the typical particle velocity $U^*$ may be obtained by balancing the viscous lubrication pressure $\mu U^* R / d^{*2}$ with the disjoining pressure $A/(6\pi d^{*3})$, which gives $U^* = A/(6\pi \mu R d^*)$. The dimensionless particle speed computed numerically is plotted in Fig.~\ref{fig:4}(a) (solid line) versus the particle-front distance and displays a maximum at a finite distance denoted $H_0$. $H_0$ is a function of the material properties, and is typically $H_0 \simeq R(1+k_e) + \mathcal{O}(d^*)$. The non-monotony of $U(H)$ can be understood with the following handwaving argument: at a distance much larger than $H_0$, the disjoining pressure is small, so that $U(H\to \infty) \rightarrow 0$. As the particle gets very close to the front, the viscous friction increases, leading to $U(H\to 0) \rightarrow 0$. Hence, $U$ has a maximum value at finite $H$.

Now that we know $U$, the distance traveled by the particle before getting engulfed can be computed by integrating the particle speed in time as 
\begin{equation}
\label{eq:interaction_length}
\ell_\mathrm{int} = \int_0^\infty U[H(t)] \,\mathrm{d}t.
\end{equation}
In the steady framework derived here, the particle velocity does not depend explicitly on time and is a function of the particle-front distance $H(t)$ only. The evolution equation of the particle-front distance is given by 
\begin{equation}
\label{eq:H}
\frac{\mathrm{d}H}{\mathrm{d}t} = U - V.
\end{equation}

The maximum repelling speed in Fig.~\ref{fig:4}(a) corresponds to the critical engulfment speed $V_\mathrm{crit}$. Indeed, if $V$ is larger than $V_\mathrm{crit}$, then $U-V$ is always negative and the particle will eventually get engulfed. Oppositely, if $V<V_\mathrm{crit}$, a steady-state solution where $U=V$ can be achieved as the particle gets closer to the front. 

Equations\,\eqref{eq:interaction_length}\,\&\,\eqref{eq:H} are evaluated for two representative cases where $k_p/k_l = 2$ (mimicking glass particles) and $k_p/k_l = 0.2$ (mimicking oil droplets), see Fig.\,\ref{fig:2}\,(b)\,\&\.(d). For both cases $R = \SI{25}{\micro \meter}$ and material properties were chosen such to match the experimental conditions.  As mentioned earlier, the Hamaker constant $A$ has been fitted to match the critical velocity $V_{\mathrm{crit}}$ (see Fig.\,\ref{fig:2}\,(c)\,\&\,(d)), where $A_{\mathrm{oil}} \sim \SI{10e-21}{\joule}$ and $A_{\mathrm{glass}} \sim \SI{10e-19}{\joule}$.
We find a good qualitative agreement between the experimental observations and theoretical predictions of the single particle-front interactions, see Fig.\,\ref{fig:2}.
We recover the three distinct cases by varying the rate of approach (see Fig.\,\ref{fig:2}\,(b)), as well as the sudden increase in the interaction length $\ell_{\mathrm{int}}$ as we approach the critical engulfment velocity (see Fig.\,\ref{fig:2}\,(d)).   

We now focus on the two limiting cases discussed in Sec.~\ref{subsec:Experiment_ell_int} and try to predict the interaction length for (i) a front speed near the engulfment speed and for (ii) fast freezing. 

\subsection*{Critical behavior near the engulfment velocity: $V \rightarrow V_\mathrm{crit}^+$}

When the front velocity is slightly larger than $V_\mathrm{crit}$, say $V = V_\mathrm{crit} + \delta V$, the particle spends a significant amount of time near the front, getting repelled (see red circles in Fig.\,\ref{fig:2}(a)). We are interested here in the limit $V \rightarrow V_\mathrm{crit}^+$ and assume $\delta V \ll V_\mathrm{crit}$. In this case, the typical time derivative of the particle-front distance is much smaller than the front velocity, \textit{i.e.} $\frac{\mathrm{d}H}{\mathrm{d}t} = U - V \sim \delta V \ll V$ when the particle is near $H \simeq H_0$. 

We aim here at obtaining an asymptotic expression for $\ell_\mathrm{int}$ for $\delta V \ll V_\mathrm{crit}$. We approximate the full $U(H)$ curve with a simpler function that correctly describe $U(H)$ in the vicinity of $H_0$. We choose a piecewise-defined function $U_0(H)$ (black dashed lines in Fig.~\ref{fig:4}(a)) that corresponds to the second order Taylor expansion of the full $U(H)$ curve near $H = H_0$ (\textit{i.e.} parabolic approximation), truncated such that $U_0$ is non negative. To be more specific, $U_0$ is defined as
\begin{equation}
U(H) \approx U_0(H) = V_\mathrm{crit} - B (H-H_0)^2, 
\quad \text{for } \quad \rvert H-H_0\rvert < \Delta H = \sqrt{\frac{V_\mathrm{crit}}{B}},
\end{equation}
where $B = - 2 U''(H) \vert_{H=H_0}$ is a system-specific positive constant that can be found numerically. With this approximation, the time evolution of the front-particle distance follows 
\begin{equation}
\label{eq:approximate_dot(H)}
\frac{\mathrm{d}H}{\mathrm{d}t} \simeq U_0-V = -\delta V - B (H-H_0)^2,
\end{equation}
which can now be integrated analytically. The reference time $t=0$ is set at the moment where particle and front start interacting, leading to the initial condition $H(0) = H_0 + \Delta H$. Eq.\,\eqref{eq:approximate_dot(H)} can be integrated analytically such that the particle-front distance follows
\begin{equation}
\label{eq:h(t)_approx}
H(t) = H_0- \delta V \tau \tan\left(\frac{t}{\tau} - \phi\right), \quad \quad \tau = \sqrt{\frac{1}{B\delta V}},
\end{equation}
where we identify a typical time scale $\tau$. The integration constant $\phi$ is found by using the initial condition, which leads to
\begin{equation}
\phi = \arctan\left(\frac{V_\mathrm{crit}}{\delta V}\right).
\end{equation}
The time $t_\mathrm{int}$ at which the interaction stops in this simplified model is when $H(t_\mathrm{int}) = H_0 - \Delta H$, which leads to 
\begin{equation}
\label{eq:t_int}
t_\mathrm{int} = 2\phi \tau \simeq \pi \tau = \pi B^{-1/2} (V-V_\mathrm{crit})^{-1/2}, 
\end{equation}
where the approximation in Eq.\eqref{eq:t_int} corresponds to the $\delta V \ll V_\mathrm{crit}$ limit. Lastly, the interaction length, or the distance travelled by the particle can be found using~\eqref{eq:interaction_length}, which leads to
\begin{equation}
\label{eq:interaction_length_asymptotic_deltaV}
\ell_\mathrm{int} = V_\mathrm{crit} t_\mathrm{int} + \mathcal{O}\left(\delta V t_\mathrm{int}\right) \simeq  \pi V_\mathrm{crit} B^{-1/2} (V-V_\mathrm{crit})^{-1/2},
\end{equation}
to leading order in the limit $\delta V \ll V_\mathrm{crit}$. Finally, we recall that the typical scale of $U(H)$ and its spatial variations are $V_\mathrm{crit}$ and $d^*$ respectively. Thus, the parameter $B= -2 U''(H) \vert_{H=H_0}$ scales as $B\sim V_\mathrm{crit}/d^{*2}$. Injecting this scaling relation into Eq.\,\eqref{eq:interaction_length_asymptotic_deltaV}, we end up with $\ell_\mathrm{int} \sim d^* (V_\mathrm{crit}/\delta V)^{1/2}$. The scaling relations of $t_{\mathrm{int}}$ and $\ell_{\mathrm{int}}$ resemble the critical exponents found for second-order phase transition models within the mean-field approximation~\cite{chaikin1995principles}. 

We show in Fig.~\ref{fig:4}(b) the trajectory earlier depicted in Fig.\,\ref{fig:2}\,(b) (solid red line), both for the full numerical solution (see red solid lines) and approximated solution Eq.~\eqref{eq:h(t)_approx} (see black dashed lines).
The approximated solution indeed provides a very good description of the full dynamics for front velocities close to the critical velocity (Fig.~\ref{fig:4}(b)). We also compute numerically the interaction length for varying front velocity as shown in Fig.~\ref{fig:2}(d). The critical behavior for $V \rightarrow V_\mathrm{crit}^+$ predicted analytically (see Eq.~\eqref{eq:interaction_length_asymptotic_deltaV}) is recovered numerically (see inset). 

We did not try to verify the prediction of the critical exponent experimentally, since it is extremely challenging to set the front velocity close enough to the critical velocity with a good enough accuracy due to intrinsic velocity fluctuations as soon as particle get in contact with the front.
This inherent variability is also responsible for the notable scatter of the experimental data (Fig.\,\ref{fig:2}(c)), as we approach this critical point and the lack of quantitative agreement between Fig.\,\ref{fig:2}(c)\&(d).
However, an excellent qualitative agreement is found, where the only adjustable parameter is the Hamaker constant, characterizing the particle-front interactions. We recall that we omitted several important contributions from \textit{e.g.} water expansion during freezing~\cite{kao2009particle}, flows driven by thermo-capillarity~\cite{park2006encapsulation}, and side wall effects (see Appendix~\ref{app:limitations}). Therefore, the fitted value of the Hamaker constant found here should not be interpreted as a precise measurement, but more as a order of magnitude. However, the interaction length criticality, meaning its scaling with $\delta V$, should not be altered by those extra physical forces. Lastly, we mentioned that there is no drastic effects of the thermal conductivity mismatch in the engulfment speed, including in the $k_e = 0$ limit.

\subsection*{Asymptotic solution at large front velocity: $V\gg V_\mathrm{crit}$}
In the fast-freezing limit where $V\gg V_\mathrm{crit}$, the particle velocity is always much smaller than the front velocity as $U$ is bounded by $V_\mathrm{crit}$. Hence, the particle-front distance rate of change can be approximated by $\frac{\mathrm{d}H}{\mathrm{d}t} = U - V \approx -V$. The particle-front distance then is approximately $H(t) \simeq H_i - Vt$, where $H_i$ is the initial condition. Therefore, the interaction length can be estimated by performing the following change of variable 
\begin{equation}
\label{eq:interaction-length_largeV}
\ell_\mathrm{int} = \int_0^{\infty} U(t) \, \mathrm{d}t = \frac{1}{V} \,\int_{-\infty}^\infty U(H) \,\mathrm{d}H.
\end{equation}
The typical scale of the particle velocity and its $H$ variations are $V_\mathrm{crit}$ and $d^*$, respectively. Therefore, the typical scaling relationship of the interaction length at large front velocity is $\ell_\mathrm{int}\sim d^*V_\mathrm{crit}/V$ (see inset of Fig.\,\ref{fig:2}(d)).

\section{Conclusion}
\label{Sec:Conclusion}

To conclude, in this paper, we have experimentally and theoretically described how two particles interact at a moving solidification front. Unidirectional freezing experiments revealed that the attractive/repulsive nature of the two-body interaction depends on the relative conductivity of the particle with respect to the surrounding liquid and does not depend on the bulk particle-particle interactions. Specifically, we find that more thermally conductive particles attract, whereas those that are less conductive repel. The extent of the particle-particle interaction is correlated to the time spent at the interface before getting engulfed in the ice. This interaction time (or length) is found to depend critically on the speed of the solidification front and therefore on the freezing rate.

We propose a model of these interactions that is inspired by the Cheerios effect for particle-particle interactions at liquid-air interfaces, but adapted to solid-liquid interfaces. 
We exploit the length scale separation of the problem between the particle-front interaction (typically $d^*$) and the perturbation of the freezing interface due to the thermal conductivity mismatch (typically $k_e R$). On the one hand, the long-range deflection of the freezing interface is entirely set by the thermal conductivity mismatch, such that particles are not facing flat but curved interfaces. An analytic asymptotic expression of the interface deflection is derived. On the other hand, the particle-front repulsion is due to particle-front intermolecular interactions and is locally determined by the small length scale $d^*$ and not affected by the second particle to leading order. However, the particles are repelled normal to the (curved) interface, causing lateral displacement and effective particle-particle interactions.

Additionally, we have explored, both experimentally and numerically, how much particles are repelled by the front prior to engulfment, meaning the interaction length.
The latter drastically increases when approaching the critical engulfment velocity. Using an asymptotic model, we find that it typically scales as $d^* (V_\mathrm{crit}/\delta V)^{1/2}$, where $d^*$ is the typical interaction length and $\delta V$ the distance of the front velocity to the critical engulfment speed. 
Finally, combining both the front deflection at larger distance and the typical interaction length close to the critical speed, we could provide theoretical predictions of the particle-particle interactions, which resemble those observed experimentally. 

The derived model paves the way towards an understanding of cluster formations during solidification. 
In this context, extending the present formalism to n-body interactions, adding effects such as solute or volume expansion, or treating freezing suspensions with different types of particles offer a wide range of perspectives.

\section*{Acknowledgements}
The authors thank Gert-Wim Bruggert and Martin Bos for the technical support and Duco van Buuren and Pallav Kant for preliminary experiments and fruitful discussions. The authors acknowledge the funding by Max Planck Center Twente, the Balzan Foundation, and the NWO VICI Grant No. 680-47-632. This work was part of JM's Ph.D. dissertation \cite{meijer2024particles}.

\appendix
\section{List of supplementary movies}

\begin{itemize}
\item \textbf{Movie 1: } Freezing of a two nearby polystyrene particles showing some effective repulsion (corresponding to Fig.\,\ref{fig:1}(a) panel I). The translational stage is moving the sample so that the freezing front appears stationary in the video. The front velocity is $V \approx \SI{0.8}{\micro \meter \per \second}$ and the video is acquired at 0.25 frames/sec and sped up 35 times.
\item \textbf{Movie 2: } Freezing of a two nearby silicone oil droplets showing some effective repulsion (corresponding to Fig.\,\ref{fig:1}(a) panel II). The translational stage is moving the sample so that the freezing front appears stationary in the video. The front velocity is $V \approx \SI{0.4}{\micro \meter \per \second}$ and the video is acquired at 0.25 frames/sec and sped up 244 times.
\item \textbf{Movie 3: } Freezing of a two nearby PMMA particles showing some effective repulsion (corresponding to Fig.\,\ref{fig:1}(a) panel III). The translational stage is moving the sample so that the freezing front appears stationary in the video. The front velocity is $V \approx \SI{1.6}{\micro \meter \per \second}$ and the video is acquired at 0.25 frames/sec and sped up 14 times.
\item \textbf{Movie 4: } Freezing of a two nearby glass particles in water showing some effective attraction (corresponding to Fig.\,\ref{fig:1}(a) panel IV). The front velocity is $V \approx \SI{0.7}{\micro \meter \per \second}$ and the video is acquired at 0.25 frames/sec and sped up 78 times.
\item \textbf{Movie 5: } Freezing of a two nearby glass particles in DMSO showing some effective attraction (corresponding to Fig.\,\ref{fig:1}(a) panel V). The front velocity is $V \approx \SI{0.14}{\micro \meter \per \second}$ and the video is acquired at 0.125 frames/sec and sped up 259 times.
\item \textbf{Movie 6: } Freezing of isolated glass particles in water for three front velocities highlighting the different scenarios depicted in Fig.\,\ref{fig:2}(a), \textit{i.e.} no particle displacement, partial rejection and full rejection. The videos are acquired at 0.25 frame/sec and sped up 115, 60 and 70 times, respectively.

\end{itemize}

\section{Numerical calculation of the front interface and particle velocity}
\label{app_sec:num}

\subsection{Non-dimensionalization}
We rescale all the lengths in \eqref{eq:front-shape} by the radius of the particle $R$ so that the dimensionless interface equation becomes
\begin{equation}
l_{\sigma} \kappa + \frac{l_{\Pi}^4}{\delta^3(\theta)} = (1+\delta)\cos(\theta)\left[1 + k_e \left(\frac{1}{1+\delta(\theta)} \right)^3 \right] - h,
\label{Eq:SI_interface_dimless}
\end{equation}
where the rescaled variables are given by  
\begin{equation}
\delta(\theta) = d(\theta)/R, \quad \quad h = H/R, \quad \quad l_{\sigma} = L_{\sigma}/R\quad \text{and} \quad l_{\Pi} = L_{\Pi}/R.
\end{equation}
The dimensionless curvature $\kappa$ follows as
\begin{equation}
\kappa =  \frac{-2(1+\delta)^2 + (1+\delta)\delta^{\prime \prime} - 3 \delta^{\prime 2}}{((1+\delta)^2 + \delta^{\prime 2})^{3/2}} + \frac{\cos \theta }{\sin \theta}\frac{\delta^{\prime}}{(1+\delta)((1+\delta)^2+\delta^{\prime 2})^{1/2}},
\label{Eq:kappa_dimless}
\end{equation}
with primes again denoting derivatives with respect to $\theta$. 

\subsection{Dynamics}
We have assumed that both the fluid transport (Stokes flow), and the heat transport are steady. Hence, equation~\eqref{Eq:SI_interface_dimless} holds for any instant $t$. We wish to determine the dynamics of the particle during the freezing. Therefore, we introduce time and set the initial particle position far from the front, typically $H(t=0) = 1.5R(1 + k_e)$. Hence, $d$, $H$ and $U$ are now time-dependent. The variation of particle-front distance is given by 
\begin{equation}
\frac{\mathrm{d}h}{\mathrm{d}t} = u(t) - v,
\end{equation}
where $u(t) = U(t)/\mathcal{W}$ and $v = V/\mathcal{W}$ are the dimensionless velocities and the time is nondimensionalized using $R/\mathcal{W}$ as a time scale. The particle velocity is set by the force balance $F_\Pi + F_\mathrm{vis} = 0$. Injecting~\eqref{Eq:SI_Fvis} and \eqref{Eq:SI_Fpi}, and using dimensionless variables, we find
\begin{equation}
u(t) \int_0^{\pi/2} \mathrm{d}\theta \sin 2 \theta  \int_\theta^{\pi/2}  \frac{\sin\hat{\theta}}{\delta^3(\hat{\theta},t)} \, \mathrm{d}\hat{\theta} + \int_0^{\pi/2} \frac{\sin 2 \theta }{\delta^3(\theta,t)}\mathrm{d}\theta = 0.
\end{equation}

\subsection{Numerical discretization}

We now provide the numerical scheme that has been used to find the particle-front dynamics in Fig.~\ref{fig:2}(b) of the main text. Let us introduce a homogeneously discretized angular axis $\theta_i = i \mathrm{d}\theta$ on the interval $[0,\theta_c]$, where $\theta_c = N\mathrm{d}\theta =  \pi/2$, with $N$ the number of angular grid points. In addition, we discretize time as $(t)^n = n \mathrm{d}t$, where superscript is used for the time discretization. The discretization of \eqref{Eq:SI_interface_dimless} is 
\begin{equation}
l_{\sigma} (\kappa)_i^{n} + \frac{l_{\Pi}^4}{ \left( \delta^3\right)_i^{n}} = (1+(\delta)_i^{n}) \cos \theta_i \left[ 1 + k_e \left( \frac{1}{1+(\delta)_i^{n}} \right)^3 \right] - (h)^{n},
\label{Eq:Equi_discrete}
\end{equation}
where the discrete curvature (see \eqref{Eq:kappa_dimless}) is given by
\begin{equation}
(\kappa)_i^{n} =  \frac{-2(1+(\delta)_i^{n})^2 + (1+(\delta)_i^{n})\left(\delta^{\prime \prime}\right)_i^{n} - 3 \left(\delta^{\prime 2}\right)_i^{n}}{((1+(\delta)_i^{n})^2 + \left( \delta^{\prime 2}\right)_i^{n})^{3/2}} + \frac{\cos \theta_i }{\sin \theta_i}\frac{\left(\delta^{\prime}\right)_i^{n}}{(1+(\delta)_i^{n})((1+(\delta)_i^{n})^2+\left(\delta^{\prime 2}\right)_i^{n})^{1/2}}. 
\end{equation}
The angular derivative are discretized by using the following finite-difference scheme
\begin{align}
\left(\delta^{\prime \prime} \right)_i^{n} & = \frac{(\delta)_{i+1}^{n} - 2 (\delta)_{i}^{n} + (\delta)_{i-1}^{n}}{\mathrm{d}\theta^2}, & \left(\delta^{\prime} \right)_i^{n} & = \frac{(\delta)_{i+1}^{n} - (\delta)_{i}^{n}}{\mathrm{d}\theta}.
\end{align}
The front-particle dynamical equation is discretized by using an explicit scheme
\begin{equation}
\label{eq:dynamics_discrete}
(h)^{n+1} - (h)^n = ((u)^n-v)\mathrm{d}t,
\end{equation}
where the particle velocity is given by
\begin{equation}
(u)^{n} \sum_{i=1}^{N} \mathrm{d}\theta \sin 2 \theta_i \sum_{j=i}^{N} \frac{\sin \theta_{N-j}}{\left(\delta^3 \right)_{N-j}^{n}}\mathrm{d}\theta + \sum_{i=1}^N \frac{\sin 2 \theta_i}{\left(\delta^3 \right)_{i}^{n}}\mathrm{d}\theta = 0.
\label{Eq:Force_discrete}
\end{equation}
We impose the following boundary conditions on the local particle-interface distance, that are
\begin{align}
\delta^{\prime}(\theta=0) = 0\quad \rightarrow \quad  (\delta)_2^{n} - (\delta)_1^{n} = 0,
\end{align}
\begin{align}
\delta(\theta = \theta_c) = \frac{h}{\cos \theta_c} -1  \quad \rightarrow \quad  (\delta)_N^{n} = \frac{(h)^{n}}{\cos \theta_N} - 1,
\end{align}
that correspond to the symmetry condition at the center and we assume an undeformed, planar interface of the solidification front far from the center. 

Assuming that $(\delta_i)^n$ and $(h)^n$ are known, the algorithm performs the following steps. First, we determine the new particle-front distance $(h)^{n+1}$ by using \eqref{eq:dynamics_discrete} and \eqref{Eq:Force_discrete}. Then, we find the new local front-particle distance solving the non-linear equation~\eqref{Eq:Equi_discrete}, using an optimised root-finding algorithm. The numerical code used in this article is available at \footnote{The numerical code can be found at \url{https://github.com/vincent-bertin/frozen_cheerios}}.

\section{Limitations: thermo-capillary, water expansion and side wall effects}
\label{app:limitations}
In the current analysis, we omitted a number of effects that should modify the particle-front interactions. Here, we briefly discuss the effects of thermocapillary migration, water expansion and side walls.

First, when placed in the thermal gradient, droplets migrate toward the warm side owing to Marangoni effects. The thermocapillary effect is well understood ~\cite{young1959motion,subramanian1983thermocapillary,meijer2023thin}, and has been already discussed in the context of droplet/freezing front interactions \cite{park2006encapsulation}. An additional force $\mathbf{F}_\mathrm{Ma}$ needs to be added in the force balance, which is given by 
\begin{equation}
F_{\mathrm{Ma}}(\theta) \sim \pi R^2 \frac{\partial \sigma}{\partial T} \int_0^{\pi/2} \mathrm{d}\theta \sin 2 \theta \int_{\theta}^{\pi/2} \frac{1}{d(\hat{\theta})} \frac{\partial T}{\partial \hat{\theta}} \mathrm{d \hat{\theta}},
\end{equation}
with $\partial \sigma/\partial T$ is the thermal coefficient of the interfacial surface tension between the droplet and the surrounding liquid. Within the time scale of the experiments ($\sim 10$ min as the typical interaction time between the oil droplet and front), the oil droplets were not observed to migrate in the liquid phase prior meeting the freezing front. This observation indicates that the thermocapillary migration velocity is far below the velocity scale arising from molecular interaction $\sim U^* \approx 1 \, \mu$m/s. Hence, we do not expect Marangoni forces to significantly affect the particle front interactions. 

Second, water expands during freezing because of the density change. As already discussed in Ref.~\cite{park2006encapsulation}, this induces an extra force $\mathbf{F}_{\mathrm{vol}}$ that reads
\begin{equation}
F_{\mathrm{vol}}(\theta) \sim -\pi \mu R^2 V \rho^{\prime} \int_0^{\pi/2} \mathrm{d}\theta \sin 2 \theta \int_{\theta}^{\pi/2} \frac{[R + d^2(\hat{\theta})] \sin \hat{\theta}}{d^3(\hat{\theta})}\mathrm{d}\hat{\theta},
\end{equation}
where $V$ is the advancing velocity of the solid-liquid interface and $\rho^{\prime} = 1 - \rho_s/\rho_l$ quantifies the change in density between the solid (ice) and liquid (water). 

Lastly, the particles and droplets studied were all buoyant and hence meet the side of the Hele-Shaw cell, while the model assumes an axisymmetric infinite system. Given that the typical particle-front interactions length scale is small compared to the particle radius, we do not expect the 3D nature of the system to fundamentally change the repulsive force. Nevertheless, possible side-wall solid-solid friction may increase the resistance to motion of the particles~\cite{saint2017interaction}. Additionally, the solid-liquid interface would deviate towards the wall, forming a meniscus~\cite{schollick2015real}. 

The addition of those effects in the model would modify the precise form of the $U(H)$ function as compared to the one shown in Fig.~\ref{fig:4}\,(a) of the main text. However, within the steady-state assumption, we anticipate that it will not modify qualitatively the non-monotonic shape of $U(H)$ drawn in Fig.~\ref{fig:4}\,(a) (see Ref.~\cite{park2006encapsulation}). Hence, the asymptotic behavior of the interaction length as function of the front velocity $\ell_\mathrm{int} (V)$ (see Fig. 2(d) of the main text) should not be altered by the water expansion and side-wall effects. We decide to ignore these effects for the sake of clarity and simplicity. 

\bibliography{References}

\end{document}